\documentclass[aps,pra,showpacs,amssymb,amsfonts,lengthcheck,twocolumn,longbibliography,superscriptaddress]{revtex4-2}

\usepackage[english]{babel}
\usepackage{amsmath}
\usepackage{float}
\usepackage{ragged2e}
\usepackage{graphicx}
\usepackage[colorlinks=true, allcolors=blue,breaklinks=true]{hyperref}
\usepackage{subcaption}
\usepackage{mathtools}
\usepackage{natbib}
\usepackage{siunitx}
\usepackage{subcaption}
\usepackage{caption}
\usepackage{xcolor}
\usepackage{balance}

\newcommand{\ket}[1]{\left\lvert #1 \right\rangle}

\DeclarePairedDelimiter{\ceil}{\lceil}{\rceil}
\DeclarePairedDelimiter\bra{\langle}{\rvert}
\DeclarePairedDelimiterX\braket[2]{\langle}{\rangle}{#1\,\delimsize\vert\,\mathopen{}#2}

\begin{document}

\title{Energetics of Rydberg-atom Quantum Computation}
\author{Óscar Alves}
\affiliation{PQI -- Portuguese Quantum Institute, Portugal}

\author{Marco Pezzutto}
\affiliation{PQI -- Portuguese Quantum Institute, Portugal}
\affiliation{Physics of Information and Quantum Technologies Group, Centro de Física e Engenharia de Materiais Avançados (CeFEMA), Portugal}
\affiliation{LaPMET -- Laboratory of Physics for Materials and Emerging Technologies, Portugal}

\author{Yasser Omar}
\affiliation{PQI -- Portuguese Quantum Institute, Portugal}
\affiliation{Physics of Information and Quantum Technologies Group, Centro de Física e Engenharia de Materiais Avançados (CeFEMA), Portugal}
\affiliation{LaPMET -- Laboratory of Physics for Materials and Emerging Technologies, Portugal}
\affiliation{Instituto Superior Técnico, Universidade de Lisboa, Portugal}
\affiliation{Quantum Green Computing, Ltd., Portugal}

\date{\today}

\begin{abstract}
    While extensive research over the past decades has been dedicated to developing scalable quantum computers, the question of their energetic performance has only gained attention more recently, but its importance is now recognized. In fact, quantum computers can only be a viable alternative if their energy cost scales favorably, and some research has shown that there is even a potential quantum energy advantage. In parallel, Rydberg atoms have recently emerged as one of the most promising platforms to implement a large-scale quantum computer. This work aims at contributing first steps to understand the energy efficiency of this platform, by investigating the energy consumption of the different elements of a Rydberg atom quantum computer. First, an experimental implementation of the Quantum Phase Estimation algorithm is analyzed, and an estimation of the energetic cost of executing it is calculated. Then, we derive an estimate of how the energy cost of performing the Quantum Fourier Transform scales with the number of qubits in the Rydberg platform. This analysis facilitates a comparison of the energy consumption of different elements within a Rydberg atom quantum computer, from the preparation of the atoms to the execution of the algorithm, and the measurement of the final state, enabling the evaluation of the energy expenditure of the Rydberg platform and the identification of potential improvements. Finally, we use the Quantum Fourier Transform as an energetic benchmark, comparing the scaling we obtained to that of the execution of the Discrete Fourier Transform in two state-of-the-art classical supercomputers. This comparison indicates that, in an ideal error-free scenario, a quantum energy advantage is achieved in the Rydberg platform for the Fourier Transform, in a regime where classical algorithms are still faster.
\end{abstract}

\maketitle

\section{Introduction}

Quantum computing represents a new paradigm in information processing~\cite{Nielsen_Chuang_2010}. By leveraging the quantum phenomena of superposition and entanglement, quantum computers may enable the execution of quantum algorithms that offer, in some cases, significant computational advantages over equivalent classical algorithms. Quantum computers also offer an advantage in the simu\-lation of physical systems where quantum effects play a significant role, such as complex many-body quantum dynamics, a task which is often computationally challenging for classical computers~\cite{feynman1982simulating, quantum_simulation}. This has applications in condensed matter physics, molecular chemistry, and materials science, among others.

Implementing quantum algorithms has proved challenging due to the difficulty of isolating and precisely manipulating quantum systems. Environmental noise and imperfect control can induce errors and decoherence, which make qubits unreliable after a certain time~\cite{BreuerPetruccione}. These factors limit the number of operations that can be performed before coherence is lost. Currently, quantum computers are in the noisy intermediate-scale quantum (NISQ) era~\cite{Preskill_2018} -- the number of qubits is small, peaking at a few hundred, and they are subject to noise, which limits their ability to perform useful algorithms. Nevertheless, much research has been conducted over the last decades in order to achieve a large-scale, fault-tolerant, quantum computer, driven by the potentially revolutionary applications of this technology. In fact, in some cases, quantum computers can solve problems that are intractable for classical computers. Shor's algorithm for factoring provides an exponential speedup over its classical counterpart ~\cite{Shor_1997} with significant implications for cryptography, as widely-used cryptographic schemes rely on the difficulty of factoring large numbers. 

Running an algorithm requires the consumption of resources, namely time, memory, and energy. We are typically interested in minimizing these resources, by designing algorithms and computation platforms that are more time-, memory- and energy-efficient~\cite{Nielsen_Chuang_2010}. Much attention has been paid to the time  efficiency of quantum algorithms, quantified by the computational complexity, as quantum advantage for many algorithms was first proven with respect to this metric. On the other hand, the question of the energy efficiency of quantum computers has only emerged more recently, driven not only by scientific interest, but also by economic and environmental concerns akin to those affecting classical computers. Indeed, only by ensuring that the energy consumption of quantum computers scales favorably, can a future large-scale quantum computer claim to be a credible real-world alternative to classical computers for selected problems. Additionally, the potential for a quantum energy advantage also exists~\cite{advantage,Pratapsi2023,energetics_trapped_ion,green,energy_initiative}, which is particularly relevant at a time when the energetic demands of classical information and communication technologies have reached unprecedented levels~\cite{masanet2020recalibrating}. For many decades the computational power of classical computers has increased steadily as described empirically by Moore's law~\cite{moore1965cramming,kish2002end}, and significant gains in energy efficiency have occurred as well~\cite{koomey2010implications}. More recently, however, the exponential growth trend has slowed, because thermal constraints and quantum effects are hindering further miniaturization~\cite{koomey2015primer}. Nowadays the energetic footprint of classical information and communication technologies accounts for an increasing amount of the world’s energy budget and is accompanied by a similar growth in their carbon footprint~\cite{gupta2022chasing}. Finally, the rise of artificial intelligence, accompanied by an increasing demand for data-center resources, is causing the expected consumption to grow even more dramatically in the next years. 

A truly comprehensive study of the energetics of quantum computation is thus paramount, addressing all its components, namely the energetic costs of the execution of the quantum gates/algorithms, of the quantum data buses, of the baseline costs of running the experimental setup (e.g., fields and lasers generating traps, vacuum, cryogenics, etc.), and of the classical control of the experiment. This research agenda should include the costs of generating non-trivial initial states, interconnecting different quantum processors, etc., and establish benchmarks to assess the energetic performance of quantum machines. Furthermore, the energetic costs will naturally depend on the chosen platform, requiring dedicated studies.

Over the years, a variety of approaches and platforms have been proposed for building quantum computers, ranging from trapped ions, to electrons in semiconductor quantum dots, single photons, superconducting circuits, Nitrogen-vacancy centers, and neutral atoms, including Rydberg atoms. While each of them presents some advantages and disadvantages, Rydberg atoms have recently attracted increasing attention~\cite{Rydberg_review}, due to the improvement in experimental techniques, which allowed the realization of better single-qubit and multi-qubit gates, control of individual atom position and displacement, and the potential for creating complex 2D and even 3D arrays of atoms \cite{Rydberg_review, CZ_gate_protocol, 256atom_array, barredo2018synthetic}. Finally, recent experiments indicate that Rydberg atoms may  allow for the leap into the fault-tolerant regime, towards implementing full quantum error correcting codes in the near future~\cite{Bluvstein2024}. 

In this work, we consider two widely known quantum algorithms, the quantum Fourier transform and the phase estimation~\cite{Nielsen_Chuang_2010}. We study their implementation on a Rydberg-atom quantum computer as provided in~\cite{phase_estimation_implementation}, and study their energetics. In Section~\ref{sec:background} we provide an overview of some key concepts, namely the two quantum algorithms, the emerging field of energetics of quantum computation -- highlighting the main developments in this new area and outlining our methodology and classification of energy costs -- and Rydberg-atom quantum computation. In Section~\ref{sec:algo-Rydberg} we review the specific implementation of the two algorithms on a Rydberg-atom quantum computer~\cite{phase_estimation_implementation}. In Section~\ref{sec:energetics} we present our estimates for the energetics of these realizations of the algorithms with Rydberg atoms, while in Section~\ref{sec:scaling} we extrapolate our findings scaling up to an arbitrary large number of qubits. The QFT was chosen for this scaling analysis due to its ubiquity as a primitive in quantum algorithms and its potential to serve as a benchmark for the energetic efficiency of quantum computation. In particular, it enables comparisons between the energetic footprints of different quantum computing platforms -- cf. related studies on trapped ions~\cite{energetics_trapped_ion}, semiconductor spin qubits~\cite{Santos2026}, and superconducting cat qubits~\cite{Ramos2026} -- as well as with classical machines. Indeed, in Section~\ref{sec:comparison}, we compare this scaling prediction to the energy cost of executing the Discrete Fourier Transform in two classical state-of-the-art supercomputers. While our comparison is just a theoretical prediction at this stage, it indicates that, above a certain input size, the quantum realization may provide an energetic advantage.

\section{Quantum algorithms, energetics and Rydberg-atom quantum computation}\label{sec:background}

\begin{figure*}[ht!]
    \centering
    \includegraphics[width=\textwidth]{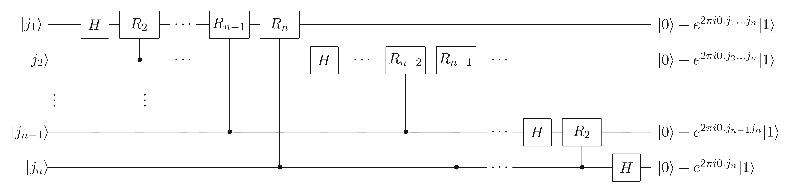}
    \caption{QFT circuit. Here $R_k =R_{z}(2\pi/2^k)$. Figure taken from~\cite{Nielsen_Chuang_2010}.}
    \label{fig:QFT}
\end{figure*}

In this section we provide an overview of some key concepts of relevance for our work. First we review the quantum algorithms which will constitute our study cases: Quantum Fourier Transform and Phase Estimation. Then we provide an overview of the emerging field of energetics of quantum computing. Finally, we provide a short introduction to Rydberg-atom quantum computation.

\subsection{Quantum Fourier Transform and Phase Estimation Algorithms}

The Quantum Fourier Transform (QFT) performs the following operation on a finite-level system basis $\ket{0}, ..., \ket{N-1}$:

\begin{equation}
    \ket{j} \rightarrow \frac{1}{\sqrt{N}}\sum_{k=0}^{N-1} e^{2\pi i jk/N}\ket{k}.
\end{equation}

The action of this operation on an arbitrary state is:

\begin{equation}
    \sum_{j=0}^{N-1}x_j\ket{j} \rightarrow \sum_{k=0}^{N-1} y_k\ket{k}, \;\; y_k=\frac{1}{\sqrt{N}}\sum_{j=0}^{N-1}x_je^{2\pi i jk/N}
\end{equation}
that is, it transforms the amplitudes, ${\{} x_j {\}}$, of the initial state into their Discrete Fourier Transform, ${\{} y_k {\}}$ \cite{Nielsen_Chuang_2010}.

The QFT is a unitary operation and, as such, can be implemented using a quantum algorithm. The circuit of this algorithm is shown in Figure~\ref{fig:QFT}.
The QFT circuit consists of the application, on the $i$-th qubit, of a Hadamard gate followed by $i-1$ controlled rotations around the $z$ axis with decreasing angles of rotation.

\begin{figure*}[ht!]
    \centering
    \includegraphics[width=0.7\textwidth]{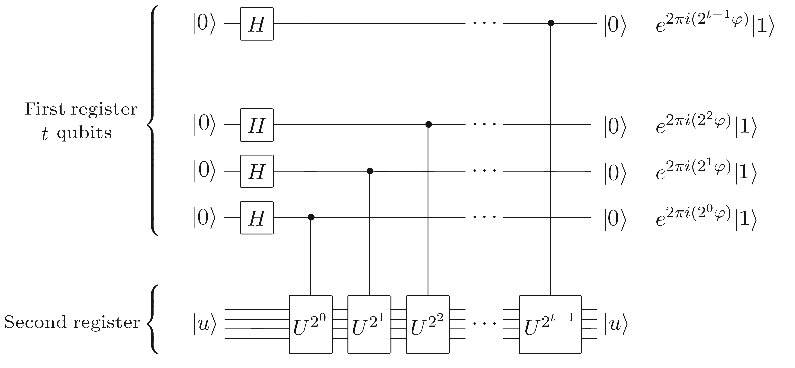}
    \caption{First part of the Phase Estimation algorithm. Figure taken from~ \cite{Nielsen_Chuang_2010}.}
    \label{fig:phaseestpt1}
\end{figure*}

The phase estimation algorithm makes use of the QFT to determine the phase of an eigenvalue of an operator $U$ when it acts on an eigenstate $\ket{u}$. Specifically, it determines $\phi$ in the expression $U\ket{u}=e^{i\phi}\ket{u}$. This algorithm uses two groups or registers of qubits: the measurement register and the phase register. The measurement regi\-ster contains $t$ qubits initially in the state $\ket{0}$. These qubits store the phase information and will be measured at the end of the algorithm. The phase register is initially in the state $\ket{u}$. The number of qubits in this register is equal to the dimension of the space where $U$ acts. This algorithm has two parts, the first of which is shown in Figure~\ref{fig:phaseestpt1} from~\cite{Nielsen_Chuang_2010}.

First, Hadamard gates are applied to every qubit in the measurement register. Then, successive controlled-$U$ operations raised to increasing powers of 2 are applied. In the second part, the inverse QFT is applied, resulting in a state that is an approximate binary value of $\phi$, which will be more exact the more qubits are used in the measurement register.

\subsection{Energetics of Quantum Computation}

In our analysis we divide the energy costs of a quantum algorithm can be divided into four types:
\begin{itemize}
    \item Baseline: energy costs of all the devices and processes needed to simply keep the quantum computer alive and idle, such as the trap generating field for trapped-ion quantum computers, the lasers that generate the optical lattice of a Rydberg-atom quantum computer, cryogenic cooling, etc.  
    \item Preparation: energy cost of all the steps required to prepare qubits to perform algorithms, such as isolation, laser cooling and qubit initialization. 
    \item Computation: energy cost of executing the successive quantum gates that comprise the algorithm.
    \item Measurement: energy cost of measuring the final state of the qubits.
\end{itemize}

Thus, the energy cost of running a quantum algorithm depends on its complexity: an algorithm that runs for a longer time, or uses more memory or performs more operations, will require more energy. However, the surrounding elements of quantum computer also require energy to function and the estimation of the energy cost of performing a quantum algorithm will have to take into account these four types of sources. Furthermore, the repetition of quantum algorithms required for acquiring sufficient statistics must also be considered, due to the inherent probabilistic nature of quantum mechanics and to the necessity of performing different quantum measurements to extract all the information contained in the quantum computer state at the end of the computation. Given the inherent sensitivity of qubits, the baseline and preparation costs are expected to be very significant across platforms.

Interest in the energy efficiency of quantum algorithms has been growing, and significant research has emerged recently. For instance, the total power consumed by Google’s Sycamore quantum processor, which claimed quantum supremacy in 2019~\cite{quantum_supremacy}, was estimated and compared to the power used by a supercomputer solving the same problem. This comparison revealed that the quantum computer's energy consumption was smaller by several orders of magnitude. Moreover, it was observed that the power consumption of the quantum processor does not change significantly between idle and running states, and it is independent of the circuit depth since most of the energy cost is attributed to refrigeration and supporting electronics (see supplementary information to~\cite{quantum_supremacy}). The energetics of a trapped-ion quantum processor was investigated in~\cite{Pratapsi2023}, which considers the realization of a classical full adder through quantum hardware, and then in~\cite{energetics_trapped_ion}, which considers the QFT performed on a 3-qubit quantum processor. The last work also contains some considerations on the scalability of the algorithm, which lead to identify a threshold for the energy advantage of a trapped-ion quantum computer performing the QFT over classical computers performing the discrete Fourier transform. More recently, the energetics of quantum gates implemented via electromagnetic waves was investigated in \cite{Stevens_2025}, showing that universal quantum computation can, in principle, be achieved with vanishing energetic cost per gate, at the expense of increased circuit complexity, while in ~\cite{stevens2026energyerrortradeoffencodingquantum} the energetic cost of Quantum Error Correction (QEC) protocols was analyzed; both works reveal a trade-off between the desired fidelity or encoding error and the energetic resources needed.

On a more fundamental level, quantum thermodynamics provides the fundamental lower bounds of energy consumption at the quantum level \cite{energy_initiative}. Taking this perspective, for instance, in~\cite{quantum_thermodynamics} the energy cost of a two-qubit photonic quantum gate was analyzed through the statistics of energy and entropy exchanges of the quantum system, when the gate is applied.

While this manuscript was in preparation, a related work \cite{vovrosh2025resourceassessmentclassicalquantum} appeared that studies the energetic consumption of a Rydberg-atom processor performing analogue quantum simulations of quantum systems. We consider this work to be largely complementary to ours, as it focuses on analog quantum simulation, whereas we address digital quantum computation.

\subsection{Rydberg atom quantum computation}
\label{sec:RydbergAtomQC}

One of the most promising recently emerged platforms for quantum computing uses neutral atoms excited to states with very high quantum numbers, known as Rydberg states~\cite{Rydberg_review}. Individual neutral atoms are isolated, confined in optical traps and cooled to low temperatures. The resulting array of atoms can then be manipulated for quantum information processing. Over the past years, this platform has experienced tremendous progress, with the control of arrays with hundreds of atoms \cite{256atom_array}, high-fidelity multi-qubit gates \cite{CZ_gate_protocol}, and entanglement of up to 20 atoms \cite{20qubit_GHZ}. Rydberg atoms have been used to implement many quantum algorithms, such as the phase estimation algorithm \cite{phase_estimation_implementation}, Grover's Algorithm \cite{grover_imp, deutsch_implementation}, the Deutsch-Josza algorithm \cite{deutsch_implementation}, and the Quantum Approximate Optimization Algorithm (QAOA) \cite{phase_estimation_implementation}. Additionally, due to their lattice arrangement and local interactions, Rydberg atoms have been particularly successful in the implementation of graph optimization problems. Examples include the Maximum Cut problem \cite{phase_estimation_implementation} and the Maximum Independent Set problem \cite{MIS_imp2, MIS_imp}.

\begin{figure}[t]
    \centering
    \includegraphics[width=6cm]{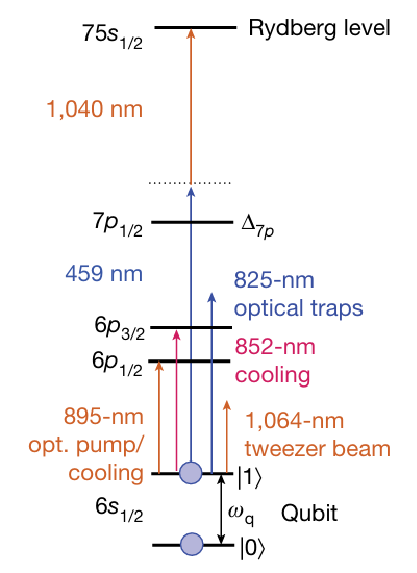}
    \caption{\justifying Atomic level diagram of the Cesium atom, with the laser wavelengths used for gate implementations, trapping and cooling. Figure taken from~\cite{phase_estimation_implementation}.}
    \label{fig:atomicdiagram}
\end{figure}

In this platform, two low energy levels of an atom are chosen as the $\ket{0}$ and $\ket{1}$ states, while an additional Rydberg level $\ket{r}$ is used to implement multi-qubit gates, as displayed in the example in Figure~\ref{fig:atomicdiagram}.  

Single-qubit gates are implemented using radiation pulses and the phenomenon of Rabi oscillations. In the presence of an oscillating electromagnetic field $\vec{E}(t) = \vec{E_0}e^{i(\omega t+\phi)}$, the Hamiltonian of an electron in an atom may be written as $H = H_0+H'(t)$, where $H_0$ is the usual atomic Hamiltonian and $H'(t)$ is the perturbation due to the electromagnetic field. The Hamiltonian that results from the interaction of the atom with radiation is: $H' = -\vec{p}\cdot \vec{E}$, where $ \vec{p}$ is the dipole moment of the atom. Thus, we can write $H'$ in the two-level system as: 
\begin{equation}
    H'(t) = \gamma e^{i(\omega t+\phi)} \ket{0}\bra{1} +\gamma e^{-i(\omega t+\phi)} \ket{1}\bra{0}
\end{equation}
where $\gamma = \bra{0}\vec{p}\ket{1}\cdot \vec{E_0}/2$.

Considering the case where the incident radiation is in resonance with the atomic transition, $\Delta = \frac{E_2-E_1}{\hbar}-\omega =0$, solving the Schrödinger equation with this Hamiltonian gives us the following matrix for the time-evolution of a two-level quantum state:
\begin{equation}
    R_\phi(\theta) = \begin{pmatrix}
\cos(\theta/2) & -ie^{i\phi}\sin(\theta/2) \\
ie^{-i\phi}\sin(\theta/2) & \cos(\theta/2)  
\end{pmatrix}
\end{equation}
where $\theta = \Omega t$ and $\Omega =\gamma/\hbar$ is the Rabi frequency. This can be used to implement any rotation of angle $\theta = \Omega t$ in the Bloch sphere around an axis in the $xy$ plane defined by the value of $\phi$.

Rotations around the $z$ axis can be implemented through the composition of 3 $xy$ rotations \cite{Rydberg_review} or through another radiation source detuned from an atomic transition that induces an AC Stark shift on the $\ket{0}$ and $\ket{1}$ states \cite{CZ_gate_protocol}. The energy shift of the states will be different and, consequently, their phases will evolve at different rates. This will generate a phase difference over time, effectively implementing a $R_z(\theta)$ gate.

\begin{figure}[h!]
    \center
    \includegraphics[width=8cm]{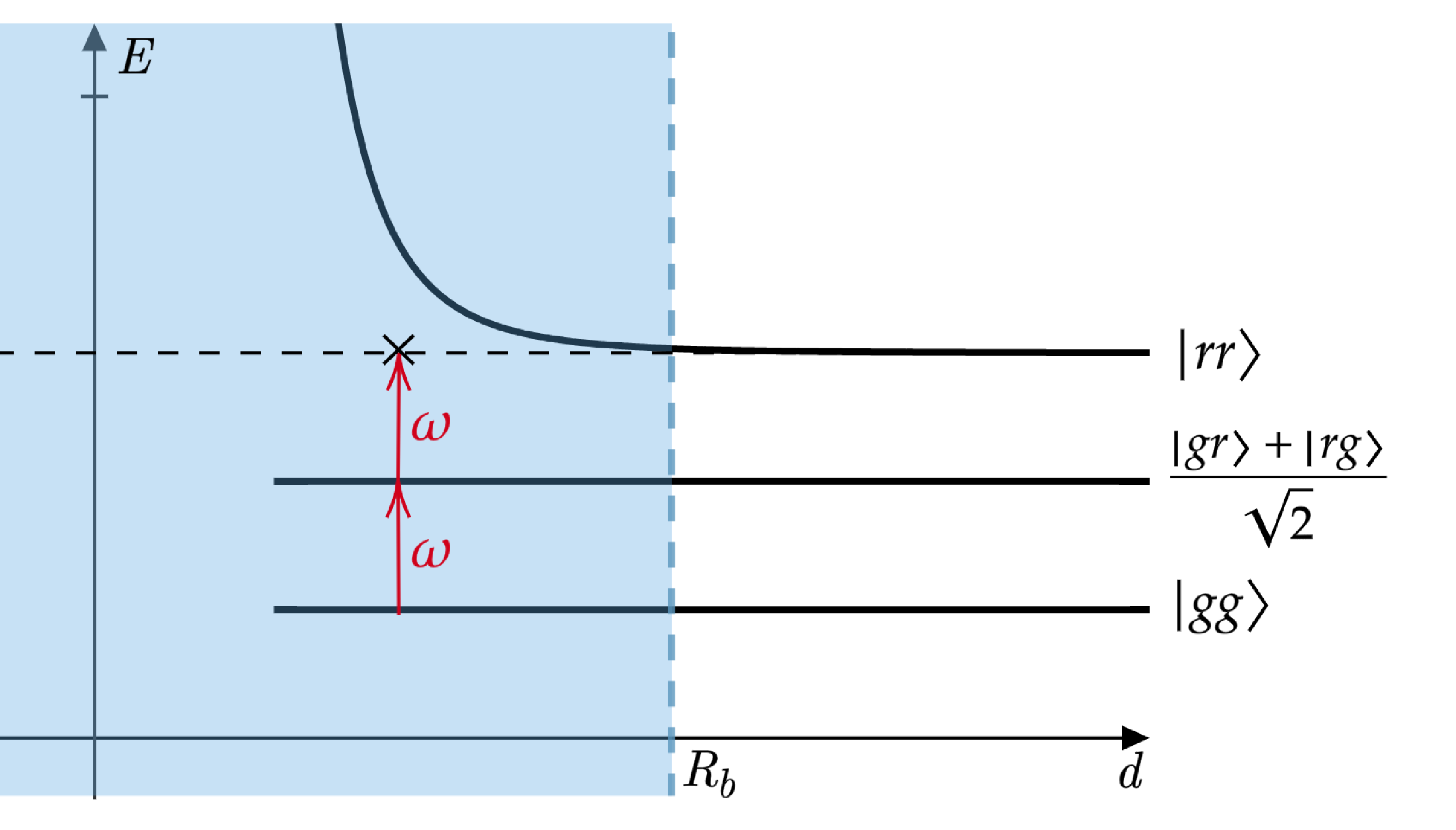}
    \caption{Rydberg blockade. When two atoms are within a distance smaller than the Rydberg radius and interact with a radiation pulse resonant with the $\ket{g} \rightarrow \ket{r}$ transition, only one atom will undergo the transition. This is because the atom that transitions to the Rydberg state will shift the energy levels of the nearby atom, causing the radiation source to no longer be in resonance with the transition of the second atom.}
    \label{fig:Rydberg_blockade}
\end{figure}

When an atom is excited to a Rydberg level, it acquires a very large dipole moment, leading to strong interactions with nearby atoms. This interaction shifts the energy levels of the surrounding atoms, meaning that radiation that previously was resonant with a certain atomic transition will no longer be resonant and will not be effective in inducing atomic transitions. This effect, known as Rydberg blockade (Figure~\ref{fig:Rydberg_blockade}) can be leveraged to implement multi-qubit gates. The energy shift will only be non-negligible within a sphere centered on the atom. The radius of this sphere is known as the Rydberg radius and can extend over several micrometers~\cite{Rydberg_review}.

\section{Quantum Fourier Transform and Phase Estimation with Rydberg atoms}
\label{sec:algo-Rydberg}

The experimental implementation of the Phase Estimation algorithm under analysis was taken from ref. \cite{phase_estimation_implementation}. In this implementation, Cesium atoms were used. The $\ket{0}$ and $\ket{1}$ states correspond to two hyperfine levels of the $6s_{1/2}$ level. The Rydberg level, $\ket{r}$, used for multi-qubit gates was $75s_{1/2}$. 

\begin{figure*}[ht!]
    \centering
    \includegraphics[width=11cm]{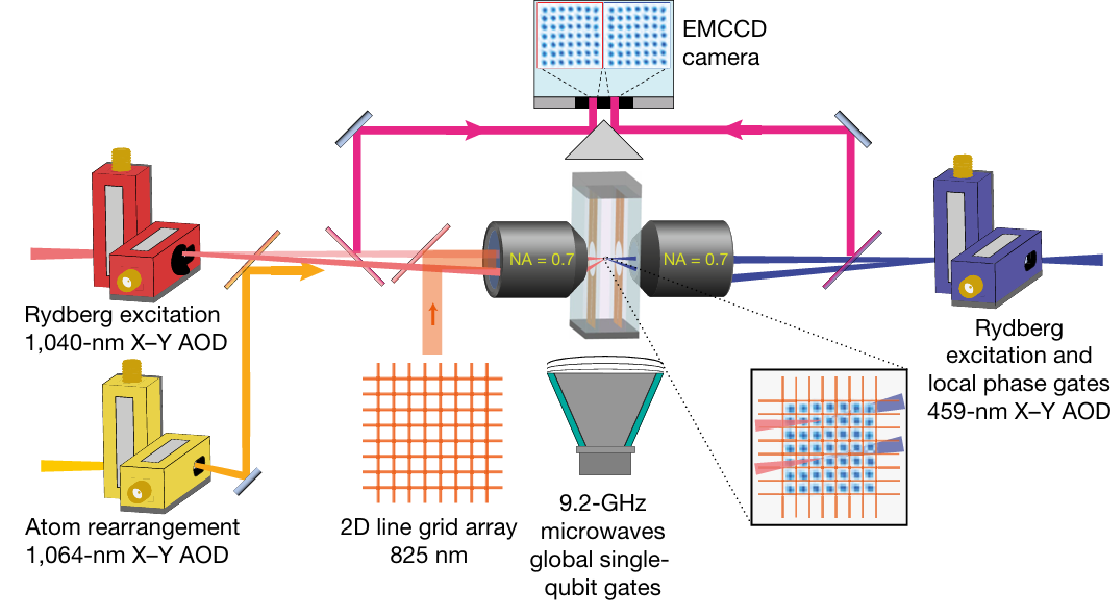}
    \caption{Experimental setup used in the implementation under analysis.~\cite{phase_estimation_implementation}}
\end{figure*}

\subsection{Qubit Preparation and Baseline}
The setup needed to prepare and execute a quantum algorithm is very complex, involving many steps and components, each with its own energy requirements. Numerous devices are required simply to keep the computer in an idle state. Therefore, analyzing the overall energy cost of these sources can become an extensive work with no clear limit on its scope. In this work, the chosen focus was on the processes that are described in ref. \cite{phase_estimation_implementation} and that directly precede computation (cooling and qubit initialization), or that are maintained throughout the entire process (optical traps), namely since the latter will scale with the number of qubits.

Throughout the preparation, execution of the algorithm, and readout, atoms are confined in optical traps. These traps are essential for positioning and holding the atoms in place, preventing undesirable interactions between them and allowing for individual addressing. They function by illuminating atoms with lasers that have spatially varying intensities. This variation creates a force that can be used to trap atoms \cite{optical_traps}. In the setup under study, blue-detuned optical traps were used, featuring beams characterized by a Gaussian radial intensity profile. To generate these traps, a laser was directed through a series of beamsplitters and other optical elements, resulting in the formation of multiple beams organized in a square grid. These intersecting beams effectively trapped atoms within the enclosed regions.

Cooling the atoms is essential to confine them in the traps, reduce decoherence of quantum states, and maintain resonance between the radiation used to address the qubits and the atomic transitions. This is achieved through laser cooling, using both Doppler and sub-Doppler methods -- such as polarization gradient cooling and gray molasses cooling -- to reach temperatures below $5\;\mu$K. Polarization cooling uses a standing wave with spatially varying polarization, creating a periodic shift in the Zeeman sublevels of an atom, which allows for a type of Sisyphus effect to occur, cooling the atoms \cite{polarization_gradient_cooling}. Gray molasses cooling uses a configuration of lasers coupled to certain atomic transitions that create a ``dark'' state where cold atoms are trapped, while hot atoms are selected to enter a cooling cycle \cite{grey_molasses}. 

After cooling, to initialize the qubits, the 895 nm laser optically pumped atoms into the $\ket{1}$ state.

\subsection{Universal Native Gate Set}
In this implementation, any quantum gate is realized through the composition of three types of native gates, which together form a universal gate set.

First, \textbf{global rotations around an axis on the xy-plane, $R_{\phi}^{(G)}(\theta)$}, are implemented using a 40 W microwave source. This source has a frequency that is resonant with the $\ket{0} \rightarrow \ket{1}$ transition, enabling the execution of the gate described in Equation (3). The Rabi frequency for this transition is $\Omega_{R_\phi^{(G)}} = 76.5 \;\si{kHz}$.
This is a global gate, meaning that, when applied, all atoms are illuminated by the microwave source and, as such, every qubit experiences the same transformation.

Second, \textbf{local rotations around the z-axis $R_{z}(\theta)$} are implemented with the 459 nm laser which is detuned by $\Delta$ from the $\ket{1} \rightarrow 7p_{1/2}$ transition, using the differential Stark shift principle described in Section~\ref{sec:RydbergAtomQC}. The differential Stark shift was: $\Omega_{R_z} = 600 \;\si{kHz}$.

Using these two gates, any local single-qubit gate can be achieved through the decomposition: $R_{\phi}(\theta) = R_{\phi+\pi/2}^{(G)}(-\alpha)R_{z}(\theta)R_{\phi+\pi/2}^{(G)}(\alpha)$. For the target qubit, these three gates collectively perform the following actions: rotate the desired axis of rotation, $\phi$, to align with the $z$ axis, execute a rotation of $\theta$ around the $z$ axis, and finally, rotate the axis of rotation back to its initial position. For all other qubits except the target qubit, only the global rotations will be applied and these will cancel each other.

Third, the native gate used to obtain multi-qubit gates was the \textbf{Controlled-Z gate}. The CZ gate protocol was taken from \cite{CZ_gate_protocol}. This gate was implemented with a two photon transition $\ket{1} \rightarrow \ket{r}$, using the 459 nm and 1040 nm lasers. To execute the gate, two pulses are consecutively emitted interacting with both qubits. The state $\ket{00}$ is uncoupled and suffers no change. States $\ket{01}$ and $\ket{10}$ oscillate between themselves and states $\ket{0r}$ and $\ket{r0}$, respectively, at a Rabi frequency $\Omega_{CZ} = 1.7 \;\si{MHz}$. After both pulses, each state returns to the initial state with an accumulated phase $\phi_{01} = \phi_{10}$. On the other hand, due to the Rydberg blockade effect, the state $\ket{11}$ oscillates between itself and $W = \frac{1}{\sqrt{2}}(\ket{1r}+\ket{r1})$ and, as a result, the Rabi frequency will be $\sqrt{2}\Omega$. After both pulses, it returns to the initial state with an accumulated phase $\phi_{11}$. The length of the pulses is chosen so that $\ket{11}$ completes one oscillation in each pulse. Both of the accumulated phases are functions of $\Omega/\Delta$. Choosing $\Delta = 0.377\Omega$ results in $\phi_{11}=2\phi_{01}-\pi$ and $\phi_{01} = 1.254$. This operation is represented by the matrix below which, when composed with local rotations, is equal to the CZ gate:

\begin{equation}
    CZ = 
(R_z(-\phi) \otimes R_z(-\phi))\begin{pmatrix}
1 & 0 & 0 & 0\\
0 & e^{i\phi} & 0 & 0\\
0 & 0 & e^{i\phi} & 0\\
0 & 0 & 0 & e^{i(2\phi-\pi)}
\end{pmatrix} .
\label{eq:CZ}
\end{equation}

\subsection{Measurement}
The measurement of the qubits has two steps. First, atoms in the $\ket{1}$ state are pushed out of the optical traps with a resonant laser beam. Then the 852 nm laser illuminates the traps, and the resulting fluorescence is measured. A dark (bright) signal indicates a quantum state of $\ket{1}$ ($\ket{0}$).

\subsection{Phase estimation}

This experimental setup was used, among other things, to implement the Quantum Phase Estimation algorithm for different operators. The implementation was divided into two parts. 

Firstly, four simple operators with well-determined phases were tested: I, Z$^{1/2}$, Z, and Z$^{3/2}$ which have phases $\phi = 0, \pi/2, \pi, 3\pi/2$ respectively. As these phases can be expressed exactly in a binary system with 2 bits, 3 qubits were used, one for the phase register and two for the measurement register. 

In the second part, the objective was to determine the molecular energy of a hydrogen molecule. To do this, the Hamiltonian of a hydrogen molecule was simulated using quantum gates $H = a_0+a_1Z+a_2X$ with  $a_0=-0.328717$ Ha, $a_1=0.787967$ Ha and $a_2=0.181289$ Ha. Then, the algorithm was used to determine the phase of the exponential of the Hamiltonian, which, when applied to an eigenstate, will be the product of the energy of this state and the time (which was a fixed parameter): $U\ket{\psi}=e^{iHt}\ket{\psi} = e^{iEt}\ket{\psi}$. Dividing the phase by $t$, the binding energy of a hydrogen molecule is obtained. This time, four qubits were used, one for the phase register and three for the measurement register. 

\section{Estimation of the energetic cost of executing the Phase Estimation algorithm}
\label{sec:energetics}

\subsection{Computation}

In this section, an estimation of the energy expenditure needed to execute the phase estimation algorithm with the previously described setup is offered. Specifically, the focus is placed on the case where the operator whose phase is to be estimated is the exponential of the Hamiltonian of the hydrogen molecule, as this is the case with more importance and relevance due to its potential applications.

To estimate the energy cost of the computation it is necessary to decompose the phase estimation circuit into the native gates. The decomposition used experimentally can be found in \cite{phase_implementation_sup}. Following the decomposed circuit, the energy cost of every gate was calculated. Since every gate is implemented using a radiation pulse, the cost will be the product of the power of the radiation source used and the pulse length. 

The power of the 459 nm laser is typically 50 mW at the atom site, with 50\% losses between the source and the atom - meaning a power of 100 mW at the source. On the other hand, the 1040 nm laser illuminates all atoms at once from the side of the register, with 10 W of power reaching the array after 10\% of losses.

For the microwave source, assuming that a cylindrical resonator was used, a power of 57.4 mW was calculated from the Rabi frequency through the expression:

\begin{equation}
    P(\omega) = \frac{1}{2}\frac{A_{\text{rad}}(\omega)}{A_{\text{dip}}}\hbar\Omega^2
\end{equation}
where we used the areas:
\begin{equation}
    A_{\text{dip}} = \frac{\mu_0}{\hbar c}\mu_B \;\;\;\text{and}\;\;\; A_{\text{rad}}(\omega) = \frac{4I_{11}}{p'_{11}}\frac{\pi a^2}{\sqrt{1-(\frac{cp'_{11}}{\omega a}})^2}.
\end{equation}
This expression was derived in \cite{Pratapsi2023} for a cylindrical cavity resonator. $p'_{11}$ is the first zero of the first Bessel function, $I_{11}$ is an integral of Bessel functions, $\omega$ is the transition frequency between the $\ket{0}$ and $\ket{1}$ states and $a$ is the corresponding cavity radius.

The pulse length needed to implement a gate can be calculated using the relation $\theta=\Omega t$. $\theta$ corresponds to the desired rotation angle for the global $xy$ rotations and the local $z$ rotations. In the CZ gate, $\theta = 2\pi$ and the Rabi frequency is $\sqrt{2}\Omega_{CZ}$, since the two pulses will have the length required to return the $\ket{11}$ to the initial state. Additionally, every time this gate occurs, these pulses are followed by two pulses that execute the local rotations described in Equation (5).

Using this process, the pulse length for every quantum gate was calculated and the total time the three radiation sources used for quantum gate implementation were turned on was determined. Multiplying by the respective power the energetic cost was obtained:

{\begin{table}[H]
\centering
\begin{tabular}{|l|l|l|l|}
\cline{1-4}
Radiation source & t{[}$\mu$s{]} & Power {[}mW{]} & $E$ {[}mJ{]}   \\ \cline{1-4}
Microwave        &   615.999     & 57.4            &  0.035                    \\ \cline{1-4}
459 nm laser      & 279.581       & 100    &   0.028                  \\ \cline{1-4}
1040 nm           & 54.454       &    $1.1 \cdot 10^4$           &    0.599                 \\ \cline{1-4}
\end{tabular}
\end{table}
}

Thus, the total energy cost of a single execution of the algorithm is $0.662\;\si{mJ}$. Considering the algorithm was repeated 700 times, the total computation energy cost was $463.4 \;\si{mJ}$.

\subsection{Baseline, preparation and measurement}

For the power of the lasers used for optical pumping and laser cooling a value of 1 mW is typically used and losses are negligible. The cooling time is typically 100 ms, while optical pumping times of 10 ms have been reported in the literature \cite{radnaev2025}.

For the optical traps, a power of 3 mW is typically needed per individual trap. However, there is a loss of 70\% of power between the source and atoms, so a power of 10 mW is needed at the source. Since in this implementation a 7$\times$7 array of traps was used, 490 mW of power was needed to supply it. These traps are considered as active during the whole optical pumping, cooling, computation and measurement steps, for their respective duration.

For the measurement, four beams were used, each with 220 $\mu$W power during 90 ms~\cite{phase_estimation_implementation}.

Once again, to calculate the energy cost, we multiply the power of these radiation sources by the time they were active. The results are summarized in the following table:

\begin{table}[H]
    \centering
    \begin{tabular}{|l|l|l|l|}
    \cline{1-4}
        Source & Power [mW] & Time [ms] & Energy [mJ]\\ \cline{1-4}
        Optical traps & 490 & $200.9$ & $98.44$\\ \cline{1-4}
        Measurement & 0.880 & 90 & 0.0792\\ \cline{1-4}
        Initialization & 1 & 10 & 0.01 \\ \cline{1-4}
        Cooling & 1 & $100$ & 0.1\\ \cline{1-4}
    \end{tabular}
    \label{}
\end{table}

Thus, considering both the execution quantum gates and the baseline, preparation and measurement costs, the total energy needed to execute the phase estimation algorithm 700 times was 69.5 J.

From these results, it is clear that the preparation and measurement stages contribute only marginally to the total energy cost, each amounting to approximately $0.1 \,\si{ m J}$. The computation cost requires a higher amount of energy -- several tenths of a mJ -- with its dominant contribution coming from the $xy$ rotations, due to the higher power of the microwave source. However, the dominant cost is clearly that of the baseline, namely the optical traps, which contribute on the order of $10^2\,\si{ m J}$.

We stress once again that this analysis does not constitute a complete inventory of all the costs associated with running an algorithm in a Rydberg atom quantum computer. In particular, in this first study, we have not included yet classical costs, namely the power consumption of the control electronics, and the cost of classical computation that always accompanies an implementation of a quantum algorithm, such as compilation and data processing. Note also that, depending on the application of the QPE, a significant energy cost could also arise from the preparation of a ground state with substantial overlap with the target state and its encoding into the qubits.

\vspace{-2mm}
\section{Scaling the energetic cost of the QFT}
\label{sec:scaling}

\vspace{-2mm}

In the $n$ qubit Quantum Fourier Transform the number of Hadamard gates increases linearly with the number of qubits, whereas the number of controlled rotations scales as $(n - 1) + (n-2) + ... + 1 = n(n-1)/2$.

The energy cost of the controlled rotations depends on the angle of rotation, which decreases. In general, the $CR_z(\pi/2^m)$ can be decomposed into native gates as:

\vspace{-2mm}
\begin{align}
        CR_z(\pi/2^m) = 
        &(I\otimes R_z({\pi}/2^m))(I\otimes H)CZ(I\otimes H) \nonumber \\ 
        &(I\otimes R_z(-\pi/2^{m+1})) (I\otimes H)CZ(I\otimes H)
        \nonumber \\ 
        &(I\otimes R_z(-\pi/2^{m+1})).     
\end{align}

Using the same procedure as in the previous section, the energy cost of one Hadamard gate can be estimated: $E_H =  4.98 \;\si{\mu J}$ and the same can be done for the CZ gate: $E_{CZ} =  47.3 \;\si{\mu J}$. Considering that $P = 100\,\si{mW}$ is the power of the 459 nm laser, the energy consumed by the $R_z(\pi/2^m)$ gate is $E_{R_z(\pi/2^m)} = \frac{P}{\Omega}\frac{\pi}{2^m}$. The energy cost of the $CR_z(\pi/2^m)$ gate is:
\begin{align}
    E_{CR_z(\pi/2^m)} &= 4E_H+2E_{CZ}+\frac{P\pi}{\Omega} \left( \frac{1}{2^m}+\frac{2}{2^{m+1}} \right) \nonumber
    \\ &= 4E_H+2E_{CZ}+\frac{P\pi}{\Omega}\frac{1}{2^{m-1}}
\end{align}
%\normalsize

Bringing all this together, the energy cost of the $n$-qubit QFT is:
%\footnotesize
\begin{align}
    E_{\text{QFT}} &= nE_H+\Sigma_{m=1}^n (n-m)E_{CR_z(\pi/2^m)} \nonumber
    \\ 
    &= (n+2n(n-1))E_H + n(n-1)E_{CZ} + \nonumber
    \\
    &\quad + \frac{P\pi}{\Omega} \, \Sigma_{m=1}^n \frac{n-m}{2^{m-1}} \nonumber
    \\ 
    &= \left( n+2n(n-1) \right) E_H + n(n-1)E_{CZ} + \nonumber
    \\
    &\quad + 4\frac{P\pi}{\Omega}(n-1+2^{-n}).
\end{align}
%\normalsize

The cooling, measurement and optical pumping of the qubits are done by illuminating all atoms with the radiation source simultaneously. As such, we can assume that, unless we have a huge amount of atoms, the energy costs of these processes will remain constant with the scaling.

The optical trap grid is formed by splitting the beam of a laser into multiple beams. As the number of traps increases with the number of qubits, if the power of the laser(s) used is constant, the power of each beam will decrease, therefore the traps will be less deep, which is not desirable. Let's assume we aim at keeping the power of each trap constant (at $P_{\text{trap}}=10\,\si{mW}$), and calculate how the power of the lasers that generate the traps would need to increase. As a lower bound for the energy, we assume that for $n$ atoms, $n$ traps are needed. Considering a square array, for $n$ atoms a $\ceil{\sqrt{n}}\times\ceil{\sqrt{n}}$ grid will be necessary. As a result, the power supply needed for a system of $n$ qubits will be: $P_{\text{array}} = \ceil{\sqrt{n}}^2 P_{\text{trap}}$.

The energy cost of the optical traps will be the product of this power by the time of cooling, optical pumping, computation and measurement. Of these, only the computation time $t_{\text{QFT}}$ increases with $n$. An expression for the computation time can be obtained by a process similar to the one used to derive $E_\text{{QFT}}$, but now considering only the pulse durations needed to implement each gate, that is, without multiplying by the radiation sources' powers:

\vspace{-2mm}
\begin{equation}
    t_{\text{QFT}} = (n+2n(n-1))t_H + n(n-1)t_{CZ} + 4\frac{\pi}{\Omega}(n-1+2^{-n})
\end{equation}
where $t_H = 25.7693 \,\si{\mu s}$ and $t_{CZ} = 12.2836 \,\si{\mu s}$ are the times needed to execute a Hadamard and CZ gate respectively.
Thus, the energy cost of the optical traps will scale as: 
\vspace{-2mm}

\begin{align}
    E_{\text{traps}} = \ceil{\sqrt{n}}^2P_{\text{trap}}(t_{\text{QFT}}(n)+t_{\text{prep}})
\end{align}

In the energy estimation of the phase estimation algorithm, the energy consumed by atom transports was ignored, as it was assumed that all 4 qubits are close to each other (within each other's Rydberg blockade radius) and, consequently, controlled gates can be performed between any two qubits without having to move them. However, with an increasing number of qubits, this will not be the case and we will need to consider the energy cost of qubit transports, and how many of them are needed to perform the algorithm.

Simulations have indicated that, when atoms are placed in a grid, the number of qubit transports increases linearly with the number of random 2-qubit controlled gates, with slope of 1.10~\cite{bernardo_thesis}. Therefore, in the QFT the number of qubit transports will increase proportionally to $n(n-1)$.

For energetic considerations, we have to take into account not only the number of transports, but also the length an atom needs to be transported, which will increase on average as the number of qubits increases. The average distance between atoms in a square grid with side $\ceil[\big]{\sqrt{n}}$ is

\begin{equation}
    D(n) = \frac{1}{\ceil{\sqrt{n}}^4}\sum_{i=1}^{\ceil{\sqrt{n}}}\sum_{j=1}^{\ceil{\sqrt{n}}}\sum_{k=1}^{\ceil{\sqrt{n}}}\sum_{l=1}^{\ceil{\sqrt{n}}}\sqrt{(k-i)^2+(l-j)^2}
\end{equation}

It can be shown that for large $n$ this expression is approximately proportional to $\sqrt{n}$.
As a result, if $E_1$ is the energy needed to transport a qubit one cell, the energy of qubit transports scales as:
\begin{equation}
    E_{\text{transport}} = 1.10 \, n(n-1)D(n)E_1.
\end{equation}

It has been shown \cite{qubit_transport_speed} that qubit transport speed needs to be lower than $0.55 \;\si{\mu m.\mu s^{-1}}$ in order to preserve quantum state fidelity. In the setup under study, beams are  separated by $3\; \si{\mu m}$ and have a width of $1 \;\si{\mu m}$. Thus, we can estimate that the time it takes to transport one atom one cell is: $t_1=4/0.55 \;\si{\mu s} =7.27\;\si{\mu s}$. Considering the power of the optical tweezers beam used to transport the atom is $P_{\text{tweezers}}=100 \, \si{mW}$ (a value that is typically used), then $E_1=P_{\text{tweezers}}t_1$. 

\begin{figure}[t]
    \centering
    \includegraphics[width=\columnwidth]{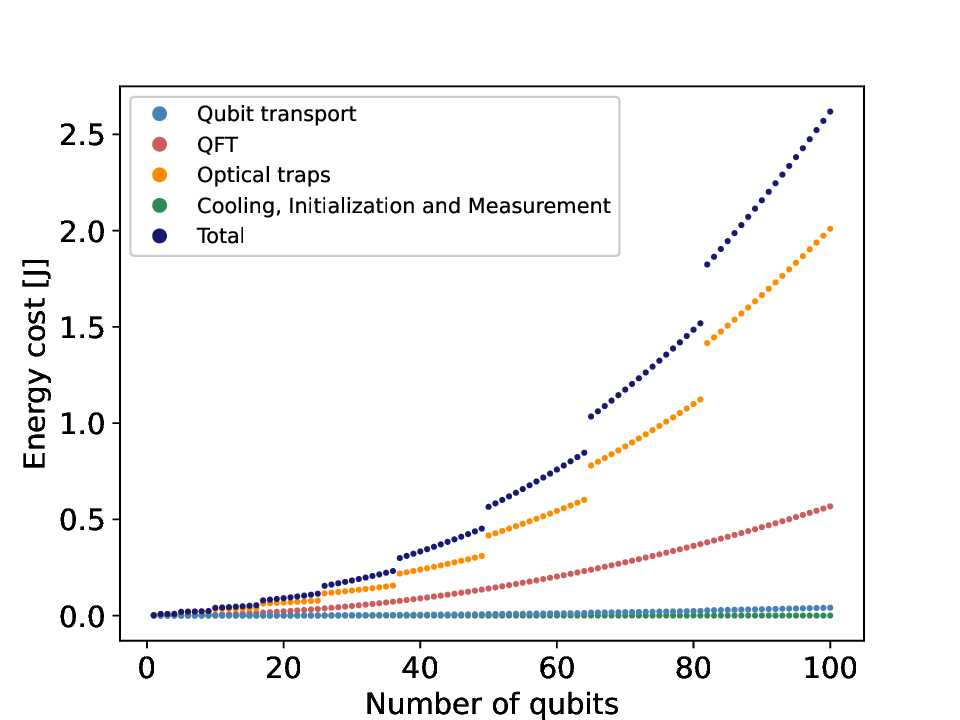}
    \caption{\justifying Scaling of the energy cost of performing the QFT in a Rydberg atom Quantum Computer, divided into different sources.}
    \label{fig:scaling}
\end{figure}

The scaling of the different sources of energy cost is shown in Figure \ref{fig:scaling}.

These expressions indicate that the energy cost of the optical traps scales the fastest, $O(n^{3})$, followed by the qubit transport, $O(n^{5/2})$ and only then the QFT gates themselves, $O(n^{2})$. Thus, we obtain a polynomial scaling of $O(n^3)$ for the total energy cost of performing the QFT in a Rydberg atom Quantum Computer. We note that these calculations consider a single run of the algorithm. However, due to the inherent probabilistic nature of quantum mechanics and to the irreversibility of quantum measurements, it is generally impossible to determine with certainty the state of a system of qubits with only a single measurement. Consequently, it is often necessary to repeat the same algorithm many times in order to reliably reconstruct the final quantum state. If the number of repetitions is assumed to be constant with the number of qubits, then our results need only be multiplied by the number of runs that is deemed necessary. This is a strong assumption, but quantum algorithms which have an advantage over classical ones are expected to satisfy it or, in the worst case, have a number of shots that increases polynomially with $n$, as they do not require a reconstruction of the full quantum system, but only measurements of local operators -- otherwise the quantum advantage would be nullified by an exponentially-increasing number of necessary shots.

We note that these scaling results relate only to the QFT algorithm itself. When the QFT is implemented to perform a practical task, these costs will be accompanied by others such as data amplitude encoding and initial state preparation. In the QPE algorithm, for instance, this would include the controlled $U$ rotations or the preparation of the initial state. The cost of these will be highly dependent on the application of the QFT that is being considered: in some cases they will be overshadowed by the QFT while in others they could scale exponentially and dominate over the QFT cost.

\section{Comparison to the Energy Consumption of Classical Computing}\label{sec:comparison}

\begin{table}[b]
\begin{tabular}{l|c|}
\cline{2-2}
& \multicolumn{1}{l|}{Energy scaling} \\ \hline
\multicolumn{1}{|l|}{Cooling, initialization and measurement} & const.                              \\ \hline
\multicolumn{1}{|l|}{QFT gates}                               & $O(n^{2})$                          \\ \hline
\multicolumn{1}{|l|}{Atom transports}                         & $O(n^{5/2})$                        \\ \hline
\multicolumn{1}{|l|}{Optical traps}                           & $O(n^{3})$                          \\ \hline
\multicolumn{1}{|l|}{Rydberg quantum computer (total)}                & $O(n^{3})$                          \\ \hline
\multicolumn{1}{|l|}{Classical supercomputers}                & $O(n2^n)$                           \\ \hline
\end{tabular}
\caption{Asymptotic energy scaling for different components of a Rydberg atom quantum computer and classical supercomputers.}
\label{tab:scaling}
\end{table}

The Quantum Fourier Transform can be used as a benchmark for the energetic efficiency of a Rydberg atom quantum computer by comparing its energetic cost to the cost of executing the Discrete Fourier Transform (DFT) on a classical computer.

To this effect, two state-of-the art supercomputers were chosen. The Top500 project \cite{top500project} ranks supercomputers based on their performance and computing efficiency when solving a system of linear equations. As of November 2025, El Capitan is the fastest supercomputer in the world, executing 1809.00 PFlop/s and having a power usage of 29685 kW. Dividing these 2 numbers, we can obtain the energy needed per flop, $1.64 \cdot 10^{-11}$ J/Flop. The most energy efficient is Jedi, for which we can do the same procedure and obtain $1.365 \cdot 10^{-11}$ J/Flop.

To compare classical and quantum Fourier Transforms we express their scaling in terms of the transform's input data size, $N$. For the QFT, this is related to the number of qubits, $n$, by $N=2^n$. A direct implementation of the DFT has a complexity of $O(N^2)$, but this can be reduced to $O(N\log_2 N)$ thanks to the Fast Fourier Transform algorithm (FFT). Consequently, the energy cost of performing a Fourier transform on classical computers grows exponentially with $\log_2(N)$, whereas for the QFT it grows polynomially. To compute the pre-factor, we follow \cite{FFT_flops} and consider that $5N\log_2N$ floating point operations are needed for $N$ samples. Multiplying this by the energy per Flop yields the scaling of the energy cost of implementing the FFT in each supercomputer

\begin{figure}[t]
    \centering
    \begin{subfigure}[b]{0.5\textwidth}
        \includegraphics[width=\textwidth]{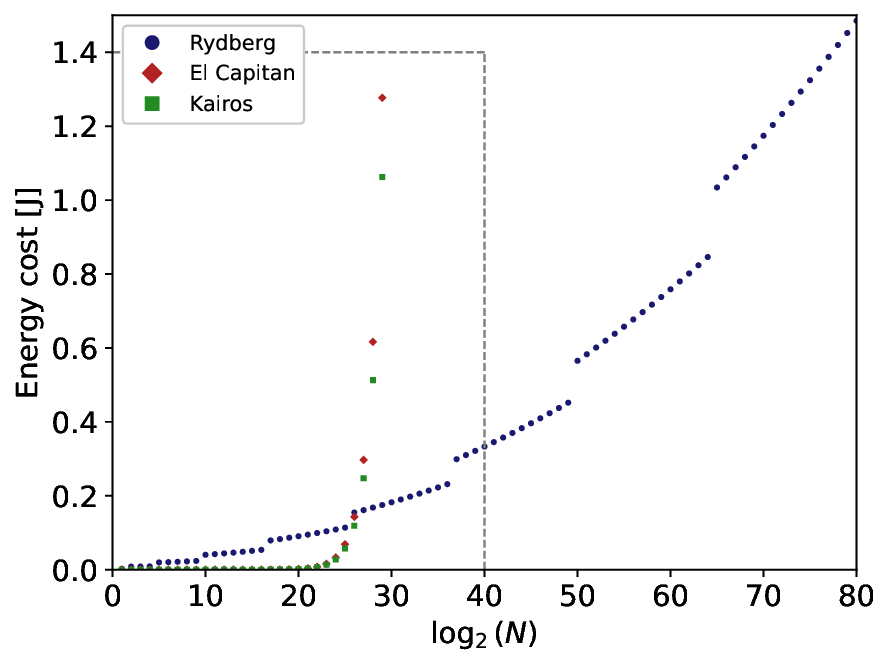}
        \caption{}
        \label{}
    \end{subfigure}
    \begin{subfigure}[b]{0.47\textwidth}
        \includegraphics[width=\textwidth]{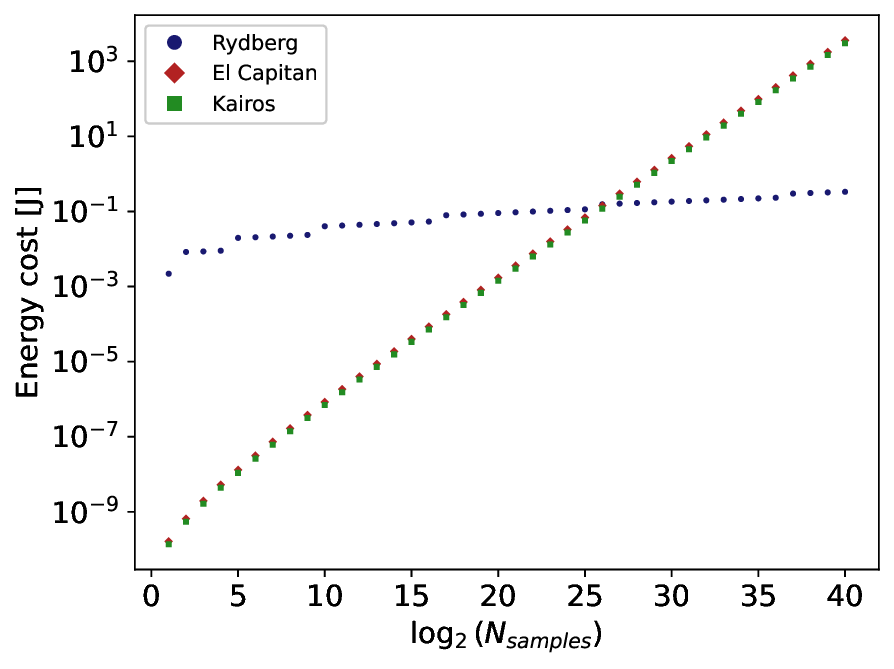}
        \caption{}
        \label{}
    \end{subfigure}
    \caption{\justifying (a) Scaling of the energy cost of performing the QFT in a Rydberg atom quantum computer as a function of the number of qubits, compared to the scaling of the cost of performing the Fast Fourier Transform in two state-of-the art classical supercomputers, El Capitan and Kairos. The scaling curves are presented as functions of the base-2 logarithm of the Fourier transform's input data size, $N$, which in the quantum case corresponds to the number of qubits. (b) A magnified view of the dashed region is shown, using a logarithmic vertical axis.}
    \label{classical_comparison}
\end{figure}

Figure~\ref{classical_comparison} presents a comparison of the scaling of the energy cost of the implementation of the quantum Fourier transform (QFT) on a Rydberg atom quantum computer, as a function of the number of qubits, with the scaling of the energy cost for performing the discrete Fourier transform (DFT) on two classical supercomputers, El Capitan and Kairos. Table \ref{tab:scaling} displays the energy cost scaling for each component of the Rydberg platform, as well as for classical supercomputers as a function of $n=\log_2(N)$.

We can see that, although for a small number of qubits the Rydberg atom quantum computer has an energy expenditure orders of magnitude higher than either supercomputer, due to the slower scaling of the energy cost in the Rydberg platform, a quantum energy advantage is achieved above 27 qubits. Even though the quantum computer has components with a significant energy overhead, their cost remains constant with the number of qubits or scales polynomially, so the exponential advantage of the quantum information processing ultimately dominates, leading to superior quantum energy performance at larger scales.

\begin{figure}[t]
    \centering
    
    \includegraphics[width=0.45\textwidth]{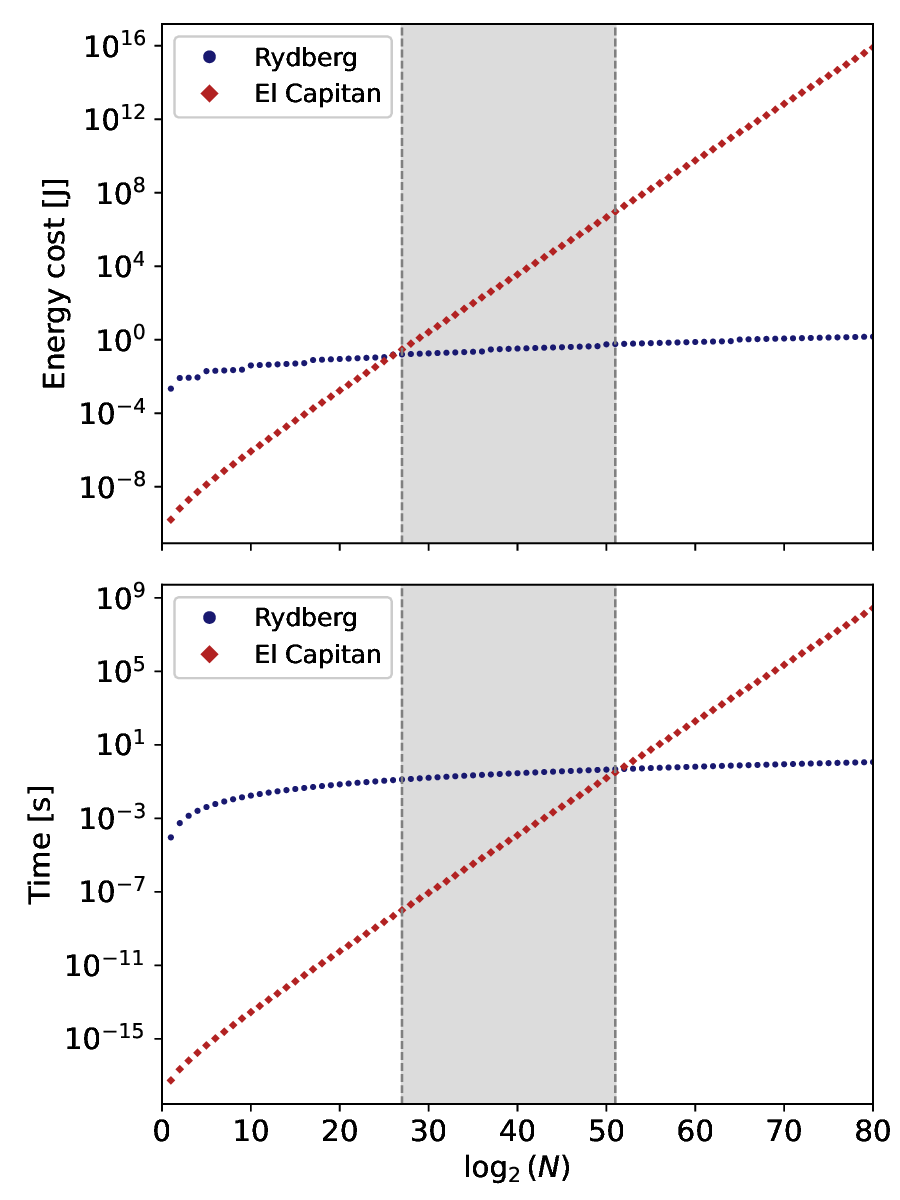}
    \caption{\justifying  Scaling of the energy cost (top) and execution time (bottom) of the QFT implemented on a Rydberg atom quantum computer, compared with the corresponding energy and time scaling of the FFT executed on El Capitan. The grey-shaded region indicates the interval where the Rydberg platform consumes less energy than the classical supercomputer while still requiring a longer execution time, i.e., the regime where there is an energy advantage without there being a computational speed-up.}
    \label{energy_vs_time}
\end{figure}

Finally, we compare the scaling of the time needed to execute the QFT and FFT, by plotting $t_\mathrm{QFT}$ (equation 11) and the classical runtime estimate: $5N\log_2 N$ multiplied by the supercomputer's speed in Flops/s. In this analysis we only consider El Capitan since it has the same energy threshold as Kairos and, being currently the fastest supercomputer, it is the most appropriate choice for a time comparison. Figure \ref{energy_vs_time} shows that --- as expected from the quantum advantage that exists for the Fourier Transform -- there exists a threshold number of qubits beyond which the Rydberg platform outperforms the fastest classical supercomputer in terms of execution time. However, this threshold (51 qubits) is significantly larger than the one obtained in the energy scaling analysis. This difference implies the existence of a region in which the quantum device consumes less energy than the classical supercomputer, before the onset of the regime where the QFT has a time advantage in this platform, suggesting that the energy advantage we observe in our estimates does not come purely from a temporal speedup. This phenomenon is a consequence of the fact that quantum gates appear to be both slower and more energy expensive than classical operations, but the difference in time is far greater than the difference in energy cost. This can be seen by comparing the ratio between the energy of a gate and energy of one flop with the ratio between the time per gate and the time per flop. For the former we obtain a value on the order of $10^{5}$, whereas for the latter we obtain a value on the order of $10^{12}$.

It is important to note that, since our scaling results do not take into account the possibility of quantum error correction, this comparison is not entirely fair for large numbers of qubits. Our scaling calculations assume an ideal error-free Rydberg atom quantum computer, but, in truth, above a certain number of qubits, the effects of errors in the algorithm will become too large and will seriously compromise the performance of the QFT. Beyond this point, while the classical procedure is more energetically costly, it will have a much higher accuracy than a noisy QFT. To obtain a reliable result with a quantum platform, it will be necessary to employ quantum error correction (QEC) protocols, with the associated overheads due to the increased number of qubits and computational steps required. However, if the error-correcting protocol used is sub-exponential in the number of qubits, we expect that the Rydberg platform will still exhibit a slower growth in energy expenditure than classical platforms, therefore still achieving a quantum energy advantage, above a certain threshold, which will be higher than the one determined here. Since the threshold for the time advantage will also be increased when accounting for QEC, we expect that the regime where there is an energy advantage without there being a time advantage will also remain. Our comparison remains valuable for a smaller number of qubits when the effect of errors is contained and there is less need for error correction.

\section{Conclusions}
In this work, we started by analyzing in depth an experimental implementation of the Quantum Phase Estimation algorithm using Rydberg atoms. We then estimated the energy cost of implementing the quantum gates required to perform this algorithm, as well as the corresponding baseline, preparation, and measurement costs. We found that the energetic costs of the computation are relatively low compared to those of the baseline, which dominates the overall energy expenditure. Thus, while implementing quantum gates is relatively cheap energetically, the surrounding components of the Rydberg atom quantum computer contribute significantly more to the total energy cost. Nevertheless, we expect that this baseline cost will be much lower than that of other quantum computing platforms, which instead of laser cooling require energetically-expensive cryogenic systems, giving the Rydberg platform a potential energy advantage over alternative quantum computing implementations.

Building on this implementation, we analyzed how the energy cost of executing the QFT would scale with the number of qubits, providing approximate expressions for the scaling of the different elements of a Rydberg atom quantum computer. The results reveal that optical traps have the steepest energy scaling, with qubit transport coming next, and only then the QFT gates themselves. As such, for a large number of qubits, the cost of maintaining the optical traps will dominate. 

Finally, we found that the scaling of the energy cost obtained for the Rydberg platform is slower than the scaling of the energy cost of the analogous Fast Fourier Transform in a classical computer, which scales exponentially faster, and identified a potential threshold for a quantum energy advantage. Remarkably, this advantage is achieved for a rather small number of qubits, already within the range accessible in current experiments. As discussed in the introduction, it has been argued that a large scale quantum computer, while theoretically outperforming classical systems, may be in practice unfeasible, due to operations requiring enormous energy and/or an unfavorable energy scaling. Our results indicate that this is not the case in the Rydberg platform: the energy cost of quantum gates is small enough for an advantage to be achieved at a small number of qubits, while the energy scaling, $O(n^3)$, differs from the complexity of the algorithm, $O(n^2)$, only by a linear factor. Moreover, we actually observe that the onset of energy advantage occurs before the regime of runtime advantage. This property ensures that in the regimes where Rydberg-atom quantum computers are useful (in the sense that they are faster than classical machines), they will always be feasible from an energetic standpoint as they will consume less energy than classical systems. 

Interestingly, this result also reveals that quantum algorithms -- originally designed with the ultimate goal of reducing runtime when compared to their classical counterparts -- are actually even better at consuming less energy. This opens the possibility for scenarios in which a Rydberg-atom quantum computer is still slower than a classical one at performing a certain task, but could nevertheless be chosen in detriment of a classical machine in order to save energy, if runtime is not the most important factor. 

This whole discussion is not limited to the QFT as the relation between the energy and time prefactors obtained in our work suggests that the same behavior (an energy advantage preceding a runtime advantage) will also occur in other algorithms with a quantum speedup. It remains an open question whether this advantage arises purely from the exponentially more compact encoding of the input data provided by quantum computing, along with the QFT's algorithmic advantage (and will disappear for algorithms where there is no computational complexity advantage), or whether it reflects a more general potential for quantum hardware to be intrinsically more energy-efficient than classical processors. Nevertheless, our results strongly indicate that in the Rydberg platform any complexity advantage will always translate to an energy advantage, achieved at number of qubits below those required for a runtime advantage.

It is important to note that these estimates assumed an ideal error-free Rydberg atom quantum computer. As natural continuation of our work, we will investigate how the fault-tolerant scenario, which requires a greater number of qubits and/or operations, will affect the energy performance. Nevertheless, since error-correcting schemes are designed to scale sub-exponentially, we expect that the energetic scaling of the Rydberg platform will remain slower than that of classical systems, and a quantum energy advantage will still be achieved. Additionally, a further study will analyze how the number of repetitions necessary to obtain a certain accuracy in the final measurement scales with the number of qubits. 

By addressing the energetic costs of the computation, qubit transport, and baseline energy expenditure, this work thus provided the first steps of our agenda for a comprehensive study of the energetics of Rydberg-atom quantum computation. The impact of the classical control of the experiment will also be the object of future work.
\balance

%%%%%%%%%%%%%%%%%%%%%%%%%%%%%%%%%%%%%%%%%%%%%%%%%%%%%%%%%%%%%%%%%%%%%

\vspace{3mm}

\begin{acknowledgements}
The authors thank Olivier Ezratty for feedback, and thank the support from FCT -- Funda\c{c}\~{a}o para a Ci\^{e}ncia e a Tecnologia (Portugal), namely through project UID/04540/2025 and contract LA/P/0095/2020, as well as from project EuRyQa – European infrastructure for Rydberg Quantum Computing (GA 101070144).

\end{acknowledgements}

\bibliography{references}

\end{document}